\documentclass[jkps,twoside,twocolumn,fleqn,showpacs,showkeys,showdoi]{revtex4}
\usepackage[dvips]{graphicx}
\usepackage{amssymb}
\usepackage{amsmath}

\begin{document}
\setcounter{page}{0}
\title[]{Nonlocal Excitations and 1/8 Singularity in Cuprates}

\author{Yoshiro \surname{Kakehashi}}
\email{yok@sci.u-ryukyu.ac.jp} 
\author{M. Atiqur R. \surname{Patoary}}
\author{Sumal \surname{Chandra}}
\affiliation{\vspace*{2mm} Department of Physics, University of the Ryukyus, 
Nishihara, Okinawa 903-0213, Japan \vspace{3mm}}
\date[]{to be published in the Journal of the
Korean Physical Society}

\begin{abstract}

Momentum-dependent excitation spectra of the two-dimensional 
Hubbard model on the
 square lattice have been investigated at zero temperature on the basis
 of the full self-consistent projection operator method in order to
 clarify nonlocal effects of electron correlations on the spectra.
 It is found that intersite antiferromagnetic correlations cause
 shadow bands and enhance the Mott-Hubbard splittings near the
 half-filling.  Furthermore nonlocal excitations are shown to move
 the critical doping concentration $\delta^{\ast}_{h}$, at which the
 singular quasiparticle peak is located just on the Fermi level, from
 $\delta^{\ast}_{h}=0.153$ (the single-site value) to 
$\delta^{\ast}_{h}=0.123$.  The latter suggests the occurance of an
 instability such as the stripe at $\delta^{\ast}_{h}=1/8$.  

\end{abstract}


\maketitle

\section{INTRODUCTION}

Low energy excitations and associated unusual behaviors of 
electrons in two-dimensional copper-oxide superconductors have 
been a question under debates in the past quarter century.  
Although theoretical studies based on the quantum Monte-Carlo 
(QMC)~\cite{grober00}, 
the exact diagonalization~\cite{dagotto94}, 
and the other methods such as the dynamical cluster approximation 
(DCA)~\cite{maier02,maier05} have clarified 
many aspects of the cuprate superconductors, the problems on the 
low energy excitations such as the non Fermi liquid behavior 
and the pseudogap state remain unresolved theoretically because of 
the limited range of the inter-site correlations and the 
limitation of the momentum and energy resolutions in the 
numerical calculations.

In order to describe long-range intersite correlations with high 
resolution in energy and momentum space, we have recently proposed a 
full self-consistent projection operator method (FSCPM) for 
single-particle excitations~\cite{kake09}.  
The theory is based on the projection 
operator technique for the retarded Green function, and makes use of an 
off-diagonal effective medium to describe the nonlocal correlations
efficiently. 
The off-diagonal matrix elements of the effective medium 
are determined to be consistent 
with those of the self-energy of the Green function.  We demonstrated 
on the basis of the half-filled Hubbard model on the simple cubic 
lattice that the long-range antiferromagnetic correlations cause
shadow bands in the low-energy region and sub-peaks of the
Mott-Hubbard bands in the strong interaction regime.

In this proceedings, we apply the theory to the two-dimensional 
Hubbard model which is the simplest model of the cuprates, and 
investigate nonlocal effects on the momentum-dependent 
single-particle excitation spectra at zero temperature.
We will demonstrate that the FSCPM describes well 
low-energy excitations of the doped two-dimensional Hubbard model 
with high resolutions and that the
nonlocal correlations can cause the shadow bands as well as the shift of
the Mott-Hubbard band splitting in the strong correlation regime.  
In particular, we suggest that
a Fermi surface instability should take place at the doping
concentration $\delta^{\ast}_{h}=0.123$.
This might correspond to the 1/8 instability of cuprates. 

In the following section, we briefly review the FSCPM theory of
single-particle excitations. 
In the theory we start from the Laplace transform of the retarded Green
function.  
Introducing an off-diagonal effective medium, 
we expand the memory function with use of the incremental 
cluster expansion~\cite{grafen97}.  
A remarkable point is that the off-diagonal 
effective medium is self-consistently taken into account up to the 
infinity so that high resolution is achieved in both energy and 
momentum.  
In Sec. III, we present numerical results of momentum dependent excitation
spectra of the doped Hubbard model.  
In the last section we summarize a conclusion of the present work.

\section{SELF-CONSISTENT PROJECTION OPERATOR APPROACH}

We consider the retarded Green function, and adopt the Hubbard 
Hamiltonian $H$ with nearest-neighbor transfer integral $t$ and 
intra-atomic Coulomb interaction $U$.  In the projection operator
method~\cite{fulde95}, 
the retarded Green function $G_{ij\sigma}(z)$ is expressed with use of 
the Laplace transformation as follows. 
\begin{eqnarray}
G_{ij\sigma}(z) = 
\Big( a^{\dagger}_{i \sigma} \  {\Big |} \ \frac{1}{z-L} \,
a^{\dagger}_{j \sigma} \Big) \ .
\label{rg1}
\end{eqnarray}
Here the inner product between two operators $A$ and $B$ is defined by  
$(A|B)=\langle [A^{+},B]_{+} \rangle$, 
$\langle \ \ \rangle$ ($[\ , \ ]_{+}$) being the thermal average 
(the anti-commutator).
$a^{\dagger}_{i\sigma}$ ($a_{i \sigma}$) is the creation (annihilation)
operator for an electron with spin $\sigma$ on site $i$. 
Furthermore $z$ denotes the complex energy variable $z=\omega+i\delta$ with 
$\delta$ being an infinitesimal positive number,
and $L$ is a Liouville operator defined by $LA=[H,A]_{-}$ for an 
operator $A$.  Here $[\ , \ ]_{-}$ is the commutator.

The Green function is expressed as follows according to the Dyson
equation. 
\begin{eqnarray}
G_{ij\sigma}(z) =
[(z - \mbox{\boldmath$H$}_{0} - 
\mbox{\boldmath$\Lambda$}(z))^{-1}]_{ij\sigma}
 \ .
\label{rg2}
\end{eqnarray}
Here $(\mbox{\boldmath$H$}_{0})_{ij\sigma}$ is the Hartree-Fock
Hamiltonian matrix and  
$(\mbox{\boldmath$\Lambda$}(z))_{ij\sigma} = \Lambda_{ij\sigma}(z) 
= U^{2} \overline{G}_{ij\sigma}(z)$ is the self-energy matrix. 
Reduced memory function $\overline{G}_{ij\sigma}(z)$ is given by
\begin{eqnarray}
\overline{G}_{ij\sigma}(z) = 
\Big( A^{\dagger}_{i \sigma} \  {\Big |} \ \frac{1}{z-\overline{L}} \,
A^{\dagger}_{j \sigma} \Big) \ .
\label{rmem}
\end{eqnarray}
The operator $A^{\dagger}_{i \sigma}$ is defined by 
$A^{\dagger}_{i \sigma}=a^{\dagger}_{i \sigma} \delta n_{i -\sigma}$
with $\delta n_{i -\sigma} = n_{i -\sigma} -  \langle n_{i -\sigma}
\rangle$, and $\langle n_{i\sigma} \rangle$
is the average electron number on site $i$ for spin $\sigma$.
$\overline{L}$ is a Liouville operator acting on the
space orthogonal to the space $\{ |a^{\dagger}_{i\sigma}) \}$;
$\overline{L}=QLQ$, $Q=1-P$, and
$P= \sum_{i\sigma} \big| a^{\dagger}_{i \sigma} \big) \, 
\big(a^{\dagger}_{i \sigma} \big|$. 

In the FSCPM theory, we introduce an energy-dependent Liouville
operator $\tilde{L}(z)$ for an effective Hamiltonian with 
an off-diagonal medium $\tilde{\Sigma}_{ij\sigma}(z)$;  
$\tilde{H}_{0}(z) = H_{0} + \sum_{ij\sigma} 
\tilde{\Sigma}_{ij\sigma}(z) \, a^{\dagger}_{i \sigma}a_{j \sigma}$.
Here $H_{0}$ is the Hartree-Fock Hamiltonian. 
The Green function 
$F_{ij\sigma}(z)$ for the Liouvillean $\tilde{L}(z)$ is expressed as 
\begin{eqnarray}
F_{ij\sigma}(z) =
[(z - \mbox{\boldmath$H$}_{0} 
- \tilde{\mbox{\boldmath$\Sigma$}}(z))^{-1}]_{ij\sigma}
 \ ,
\label{fij}
\end{eqnarray}
where $(\tilde{\mbox{\boldmath$\Sigma$}}(z))_{ij\sigma} = 
\tilde{\Sigma}_{ij\sigma}(z)$.
It should be noted that the Green function $F_{ij\sigma}(z)$ becomes
identical with $G_{ij\sigma}(z)$ when 
\begin{eqnarray}
\tilde{\Sigma}_{ij\sigma}(z) = \Lambda_{ij\sigma}(z) \ .
\label{sceqij}
\end{eqnarray}

In order to obtain an explicit expression of the self-energy 
$\Lambda_{ij\sigma}(z)$,
we separate the Liouvillean $L$ into $\tilde{L}(z)$ and the
remaining interaction part $L_{\rm I}(z)$, {\it i.e.}, 
$L = \tilde{L}(z) + L_{\rm I}(z)$, and
expand the memory function (\ref{rmem}) with respect to the interaction
by using the incremental method~\cite{grafen97}.  
In the present calculations, we take into account the intersite
correlations within the pair-site approximation; 
\begin{eqnarray}
\overline{G}_{ii \sigma}(z) &=&
\overline{G}^{(i)}_{ii \sigma}(z) + 
\sum_{l \neq i} \Delta \overline{G}^{(il)}_{ii \sigma}(z) \ , 
\label{icrii}  \\
\overline{G}_{ij \sigma}(z) &=&
\overline{G}^{(ij)}_{ij \sigma}(z) \ .
\label{icrij}
\end{eqnarray}
Here $\Delta\overline{G}^{(il)}_{ii\sigma}(z) = 
\overline{G}^{(il)}_{ii\sigma}(z) - \overline{G}^{(i)}_{ii\sigma}(z)$.  
These terms are calculated from cluster memory functions defined by 
$\overline{G}^{({\rm c})}_{ij\sigma}(z) =  
\big( A^{\dagger}_{i\sigma} \  {\big |} 
(z-\overline{L}^{({\rm c})}(z))^{-1} \,
A^{\dagger}_{j\sigma} \big)$ $(c= i, ij)$.
$\overline{L}^{({\rm c})}(z)=QL^{({\rm c})}(z)Q$ and 
$L^{({\rm c})}(z)$ is the Liouvillean for a cluster
$c$ embedded in the off-diagonal medium 
$\{ \tilde{\Sigma}_{lm\sigma}(z) \}$.

In the calculation of the cluster memory functions mentioned above, 
we applied the form obtained by the renormalized 
perturbation theory~\cite{kake04-2}.
\begin{eqnarray}
\overline{G}^{({\rm c})}_{ij\sigma}(z) = 
\Big[ \overline{\mbox{\boldmath$G$}}^{({\rm c})}_{0}(z) \cdot 
(1-\overline{\mbox{\boldmath$L$}}^{({\rm c})}_{I} \cdot 
\overline{\mbox{\boldmath$G$}}^{({\rm c})}_{0}(z))^{-1} \Big]_{ij\sigma} \ .
\label{memcij}
\end{eqnarray}
Here $(\overline{\mbox{\boldmath$L$}}^{({\rm c})}_{I})_{i\sigma
j\sigma^{\prime}} = U (1- 2 \langle n_{i -\sigma}
\rangle)/\chi_{i\sigma}$, and
$\chi_{i\sigma} = \langle n_{i-\sigma} \rangle 
(1 - \langle n_{i-\sigma} \rangle)$. 
In the simplest approximation~\cite{kake04-2}, 
$(\overline{\mbox{\boldmath$G$}}^{({\rm c})}_{0})_{ij\sigma}(z)$ is given by
\begin{eqnarray}
(\overline{\mbox{\boldmath$G$}}^{({\rm c})}_{0})_{ij\sigma}(z) 
&=& \hspace{3mm} \nonumber \\ 
& & \hspace*{-25mm} A_{ij\sigma}
\int \frac{\displaystyle
d\epsilon d\epsilon^{\prime} d\epsilon^{\prime\prime} 
\rho^{({\rm c})}_{ij\sigma}(\epsilon)
\rho^{({\rm c})}_{ij-\sigma}(\epsilon^{\prime})
\rho^{({\rm c})}_{ji-\sigma}(\epsilon^{\prime\prime})
\chi(\epsilon, \epsilon^{\prime},
\epsilon^{\prime\prime})
}
{\displaystyle 
z - \epsilon - \epsilon^{\prime} + \epsilon^{\prime\prime}}
\ . \ \ \ \ \ 
\label{rpt00}
\end{eqnarray}
Here 
$A_{ij\sigma} = [ \chi_{i\sigma} /
\langle n_{i -\sigma} \rangle_{\rm c} 
(1 - \langle n_{i -\sigma} \rangle_{\rm c})] \delta_{ij} + 1 - \delta_{ij}$,
$\chi(\epsilon, \epsilon^{\prime}, \epsilon^{\prime\prime}) =
(1-f(\epsilon))(1-f(\epsilon^{\prime})) f(\epsilon^{\prime\prime})
+ f(\epsilon)f(\epsilon^{\prime})
(1-f(\epsilon^{\prime\prime}))$, 
$f(\epsilon)$ is the Fermi distribution function, 
$\langle n_{i\sigma} \rangle_{\rm c}$ is the electron number for a
cavity state defined by $\langle n_{i\sigma} \rangle_{\rm c} = \int d\epsilon 
\rho^{({\rm c})}_{ii\sigma}(\epsilon) f(\epsilon)$, and 
$\rho^{({\rm c})}_{ij\sigma}(\epsilon) = - \pi^{-1} {\rm Im} [(
\mbox{\boldmath$F$}_{\rm c}(z)^{-1}
+ \tilde{\mbox{\boldmath$\Sigma$}}^{({\rm c})}(z))^{-1}]_{ij\sigma}$.
The coherent cluster Green function 
$(\mbox{\boldmath$F$}_{\rm c}(z))_{ij\sigma}=F_{ij\sigma}(z)$ is given by 
Eq. (\ref{fij}), and 
$(\tilde{\mbox{\boldmath$\Sigma$}}^{({\rm c})}(z))_{ij\sigma}=
\tilde{\Sigma}_{ij\sigma}(z)$ for sites $(i,j)$ belonging to the cluster $c$.
Note that a ``cluster'' $c=(ij)$ does not mean that sites
$(i,j)$ are nearest neighbors.  Instead they may be far apart.

Since we have truncated the higher-order terms in the expansions  
(\ref{icrii}) and (\ref{icrij}), the self-energy 
$\Lambda_{ij\sigma}(z)=U^{2}\overline{G}_{ij\sigma}(z)$ depends on the
medium $\tilde{\Sigma}_{ij\sigma}(z)$.  
We determine the medium self-consistently from condition (\ref{sceqij}).  
Note that the present theory reduces to the projection operator CPA 
in infinite dimensions~\cite{kake04-1}, which is equivalent to
the dynamical mean-field theory~\cite{hirooka77, georges96, kake04-3}.

The momentum dependent excitation spectra are calculated from
the Green function
\begin{eqnarray}
G_{k\sigma}(z) = \frac{1}{z - \epsilon_{k\sigma} - 
\Lambda_{k\sigma}(z)} \ .
\label{gk}
\end{eqnarray}
Here $\epsilon_{k\sigma}$ is the Hartree-Fock one-electron energy 
eigen value, 
and the momentum-dependent self-energy is calculated via Fourier
transform of the off-diagonal self-energy as
$\Lambda_{k\sigma}(z) = \sum_{j} \Lambda_{j0\sigma}(z) 
\exp (i\mbox{\boldmath$k$}\cdot\mbox{\boldmath$R$}_{j})$.
Note that the theory yields the spectra with high resolution in both
energy and momentum because it is based on the retarded Green function
and the off-diagonal effective medium is taken into account up to
infinity in distance.  In the DCA, the spectra with high resolution are
not obtained at low temperatures because it relies on the numerical
analytic continuation and a small size of cluster embedded in an
effective medium.

\section{Numerical Results}

We have performed the self-consistent calculations of excitation spectra
for the two-dimensional Hubbard model 
on the square lattice in the nonmagnetic
state at zero temperature.  In the calculations of the memory functions 
(\ref{rpt00}), we adopted the Laplace transform, 
and calculated off-diagonal self-energies up to the 50th nearest 
neighbors self-consistently.
%
%
\begin{figure}[t!]
\includegraphics[width=8.9cm]{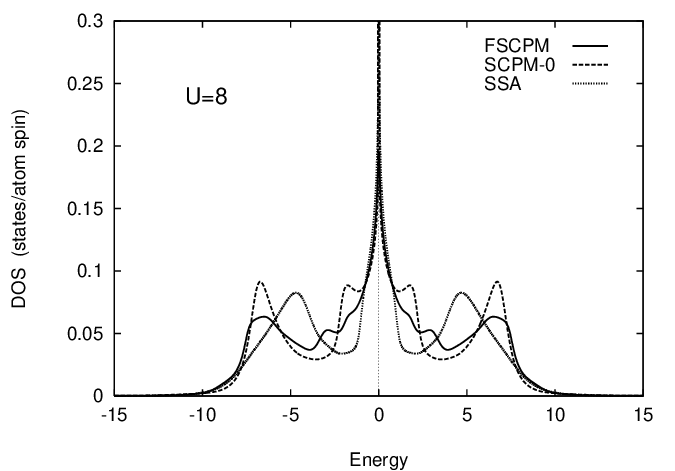}
\label{fig-dosn1.0u8}
\caption[0]{
The single-particle DOS at half-filling in the FSCPM (solid curve), 
the SCPM-0 (dashed curve), and the SSA (dotted curve).
}
\end{figure}
%
%

Figure 1 shows calculated single-particle densities of states (DOS) for
$U=8$ in unit of $|t|=1$ at half-filling.  
Note that the Mermin-Wagner theorem~\cite{mermin66} 
tells us the antiferromagnetic state with the N\'{e}el temperature 
$T_{\rm N}=0$ K at half-filling, 
while we assumed here the nonmagnetic state.  Thus the
results presented here are not directly applicable at half-filling, but
express the basic behavior of excitations in the vicinity of
half-filling.
The DOS is compared with
those of the single-site approximation (SSA) and the diagonal self-consistent
approximation (SCPM-0) in which the off-diagonal
effective medium has been neglected, but the off-diagonal self-energies 
$\Lambda_{ij\sigma}(z)$ have been taken into account.  
The result indicates that the van Hove singularity remains near the
Fermi level in the vicinity of half-filling.
We find that 
nonlocal correlations increase the splitting between the upper and lower
Hubbard bands as compared with the SSA.
On the other hand, the peaks of the both Hubbard bands are suppressed as
compared with those obtained by the SSA and the SCPM-0.
The increment of the splitting is explained by strong
antiferromagnetic correlations.  In fact the latters cause the splitting
$U+2z_{\rm NN}|J|$ instead of $U$ in the strong correlation regime.
Here $z_{\rm NN}$ is the number of the nearest neighbors, and $J$ is the
super-exchange interaction $J=-4|t|^{2}/U$.  The formula yields the
splitting 12, and explains the splitting in Fig. 1.

Nonlocal correlations also cause small peaks at energy 
$|\omega| \approx 3.0$.  This is interpreted as shadow bands of type 
$\epsilon^{\rm SDW}_{\pm}(k) = \pm 
\sqrt{\tilde{\epsilon}_{k}^{2} + \Delta^{2}}$
due to long-range antiferromagnetic correlations.  
Here $\tilde{\epsilon}_{k}$ is a quasiparticle band, $\Delta$ is an
exchange splitting.
%
%
\begin{figure}[t!]
\includegraphics[width=8.9cm]{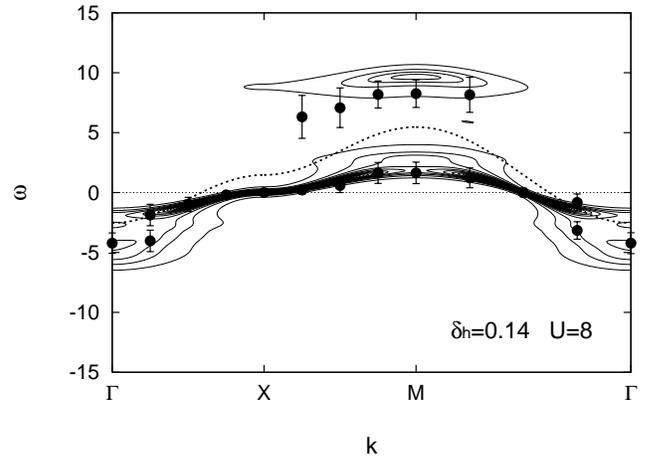}
\label{fig-dctm86}
\caption[0]{
Contour map of the momentum-dependent excitations along the
 high-symmetry lines at $\delta_{h}=0.14$.  
Closed circles with error bars are the QMC
 results at $T=0.33$.  Dashed curve shows the Hartree-Fock quasiparticle
 dispersion. 
}
\end{figure}
%
%

When holes are doped, the spectral weight of the lower Hubbard band
moves to the quasiparticle band as well as the upper Hubbard band.  
Furthermore the
upper Hubbard band shifts to the higher energy region.
Figure 2 shows an example of the momentum dependent excitation spectra 
in the optimum doped region.  The excitations of the lower
Hubbard band are damped, and those of the upper Hubbard band are
enhanced.  These results of spectra are consistent with those of the 
QMC~\cite{grober00}.  
Especially, the quasiparticle band near the Fermi level shows
a good agreement with that of the QMC.
%
%
\begin{figure}[t!]
\includegraphics[width=7.0cm]{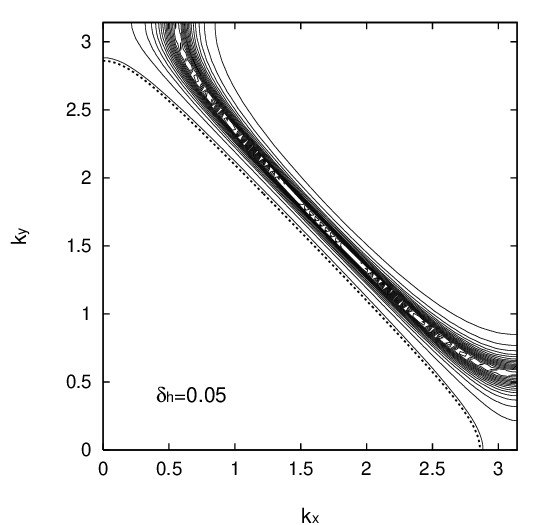}
\label{fig-cntrk95}
\caption[0]{
Excitation spectrum at the Fermi energy for $\delta_{h}=0.05$.
The dotted curve shows the Fermi surface in the Hartree-Fock
 approximation. 
}
\end{figure}
%
%
\begin{figure}[t!]
\includegraphics[width=7.0cm]{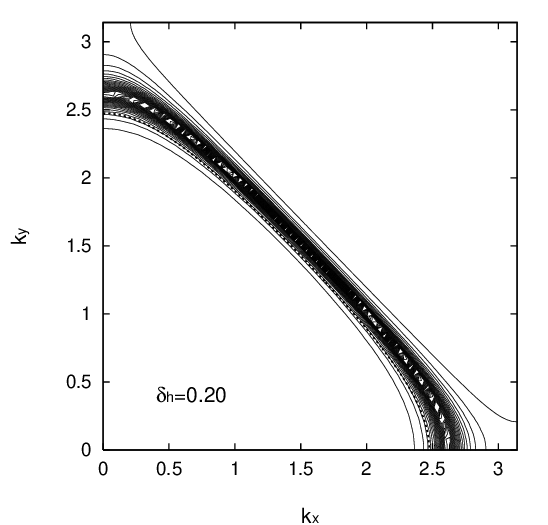}
\label{fig-cntrk80}
\caption[0]{
Excitation spectrum at the Fermi energy for $\delta_{h}=0.20$.
The dotted curve shows the Fermi surface in the Hartree-Fock
 approximation. 
}
\end{figure}
%
%

Excitations at energy $\omega=0$ determine the Fermi surface.  
We present in Figs. 3 and 4 the spectra at zero energy in the underdoped
region ($\delta_{h}=0.05$) as well as the overdoped region 
($\delta_{h}=0.20$).  Due to a strong
damping of the lower Hubbard band, a hole-like Fermi surface appears 
in the underdoped region, while in the overdoped region an 
electron-like Fermi surface appears.  
The results are in good agreement with those
based on the QMC combined with the DCA~\cite{maier02}.

Although the results of the FSCPM are consistent with those
expected from the QMC+DCA at finite temperatures, there are a few points
showing the discrepancy.  The marginal Fermi 
liquid behavior is one of them.  The marginal Fermi liquid 
is characterized by the
imaginary part of the self-energy being proportional to 
max$(|\omega|, T)$~\cite{varma89}.
In the recent $4 \times 4$ QMC+DCA calculations~\cite{vidhya09}, 
the marginal Fermi liquid behaviors are
found at finite temperatures.  The result for $U=6$ suggests that
the marginal Fermi liquid remains even at zero temperature 
at doping concentration
$\delta^{\ast}_{h}=0.15$, where the peak of the single-particle DOS is
just on the Fermi level.
This is rather close to the result of the SSA, $\delta^{\ast}_{h}=0.153$
in our calculation ({\it i.e.}, the projection operator CPA).
%
%
\begin{figure}[t!]
\includegraphics[width=8.9cm]{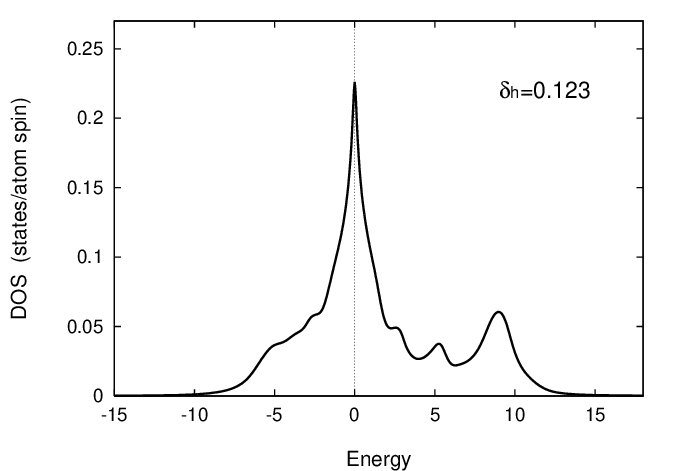}
\label{fig-dosn0.877}
\caption[0]{
The density of states at $\delta_{h}=0.123$ where the
 quasiparticle peak is just on the Fermi level.
}
\end{figure}
%
%

We point out that the peak at $\delta^{\ast}_{h}$ is connected to that
in Fig. 1 at zero doping.
With the hole doping, the peak with van Hove singularity sinks first
below the Fermi level because of the transfer of the spectral weight
from the lower Hubbard band to the upper Hubbard band.  But, it
gradually approach to the Fermi level with further doping, and the peak
is again located on the Fermi level at $\delta^{\ast}_{h}$. 

In the present calculations of the FSCPM, 
the critical doping concentration at which
the quasiparticle peak is located just on the Fermi level is 
$\delta^{\ast}_{h}=0.123$ for $U=8$ as shown in Fig. 5.  
Furthermore electrons are the
Fermi liquid instead of the marginal Fermi liquid at $\delta^{\ast}_{h}$ 
because we find that the imaginary part of the momentum 
dependent self-energy in the quasiparticle energy region is 
proportional to $\omega^{2}$ instead of $|\omega|$.
It is remarkable that the critical concentration
$\delta^{\ast}_{h}=0.123$ is close to the 1/8 stripe
instability found in the experiments~\cite{mooden88, wu11}.
Though the present results indicate the Fermi liquid behavior,  it is
possible that the Fermi surface causes an instability at
$\delta^{\ast}_{h}$ because of the peak just on the Fermi level, so that
the normal state may change to the stripe phase at 
$\delta^{\ast}_{h}=1/8$.

\section{CONCLUSION}

We have investigated the nonlocal effects of electron correlations on
the excitation spectra of the two-dimensional Hubbard model 
at zero temperature on
the basis of the full self-consistent projection operator method
(FSCPM).  The FSCPM takes into account the
long-range intersite correlations in a self-consistent way making use of
the off-diagonal effective medium, and yields the spectra with high
energy and momentum resolution.
In the strong Coulomb interaction regime ({\it i.e.}, the case of
$U=8$), we found
that the long-range intersite antiferromagnetic correlations create the
shadow band excitations.  The strong antiferromagnetic correlations also
enhance the Mott-Hubbard splitting by about $2z_{\rm NN}|J|$ where
$z_{\rm NN}$ ($J$) is the number of the nearest neighbors 
(the super-exchange interactions energy).  By comparing with the QMC 
and the QMC+DCA results, 
we verified that the FSCPM quantitatively describe the
momentum dependent excitation spectra in the doped region.  We also
found that the nonlocal correlations shift the critical concentration
$\delta^{\ast}_{h}$ at which the quasiparticle peak is located just on
the Fermi level from $\delta^{\ast}_{h}=0.153$ (the SSA) to 
$\delta^{\ast}_{h}=0.123$, 
while the QMC+DCA at finite temperatures suggest $\delta^{\ast}_{h} =
0.15$ at zero temperature.  
Furthermore the electrons at $\delta^{\ast}_{h} = 0.123$ obtained by
the FSCPM are the Fermi liquid, though the QMC+DCA suggested the
marginal Fermi liquid even at zero temperature.  
It is plausible that the Fermi surface
instability takes place because of the singular peak on the Fermi
level.  We speculate that such an anomaly corresponds to the 1/8 stripe
instability and that it may cause the non Fermi liquid behaviors at
finite temperatures as found in the QMC+DCA 
in the vicinity of $\delta^{\ast}_{h}$. 
Details of the DOS and $\delta^{\ast}_{h}$ as a function of the Coulomb
interaction $U$ will be published elsewhere.  Theoretical calculations
for possible instabilities at $\delta^{\ast}_{h}$ are left for future
investigations.

\begin{acknowledgments}

The authors would like to express their sincere thanks to Prof. Peter Fulde
 for valuable discussions.
This work is supported by a Grant-in-Aid for Scientific Research
 (22540395). 
\end{acknowledgments}

%


\begin{references}
%
%
\bibitem{grober00}
C. Gr\"{o}ber, R. Eder, and W. Hanke, Phys. Rev. B {\bf 62}, 4336 (2000).

\bibitem{dagotto94}
E. Dagotto, Rev. Mod. Phys. {\bf 66}, 763 (1994).

\bibitem{maier02}
Th. A. Maier, Th. Pruschke, and M. Jarrell, Phys. Rev. B {\bf 66},
 075102 (2002). 

\bibitem{maier05}
T. Maier, M. Jarrell, T. Pruscke, and M.H. Hettler, Rev. Mod. Phys. {\bf
 77}, 1027 (2005).

\bibitem{kake09}
Y. Kakehashi, T. Nakamura, and P. Fulde, J. Phys. Soc. Jpn. {\bf 78},
 124710 (2009).

\bibitem{grafen97}
J. Gr\"{a}fenstein, H. Stoll, and P. Fulde, Phys. Rev. B {\bf 55},
 13588 (1997).

\bibitem{fulde95}
See for example, P. Fulde, {\it Electron Correlations in Molecules and
 Solids} (Springer, Berlin, 1995).

\bibitem{kake04-2}
Y. Kakehashi and P. Fulde, Phys. Rev. B {\bf 70}, 195102 (2004). 

\bibitem{kake04-1}
Y. Kakehashi and P. Fulde, Phys. Rev. B {\bf 69}, 045101 (2004). 

\bibitem{hirooka77}
S. Hirooka and M. Shimizu, J. Phys. Soc. Jpn. {\bf 43}, 70 (1977).

\bibitem{georges96}
A. Georges, G. Kotliar, W. Krauth, and M.J. Rosenberg,
 Rev. Mod. Phys. {\bf 68}, 13 (1996).

\bibitem{kake04-3}
Y. Kakehashi, Adv. Phys. {\bf 53}, 497 (2004). 

\bibitem{mermin66}
N.D. Mermin and H. Wagner, Phys. Rev. Lett {\bf 17}, 1133 (1966); 
D.K. Gohsh, Phys. Rev. Lett. {\bf 27}, 1584 (1971) [Errata; {\bf 28},
 3301 (1972)].

\bibitem{varma89}
C. M. Varma, P. B. Littlewood, S. Schmitt-Rink, E. Abrahams, and
 A. E. Ruckenstein, Phys. Rev. Lett. {\bf 63}, 1996 (1989); 
E. Abrahams and C. M.  Varma, Phys. Rev. B {\bf 68}, 094502 (2003).

\bibitem{vidhya09}
N. S. Vidhyadhiraja, A. Macridin, C. Sen, M. Jarrell, and Michael Ma, 
Phys. Rev. Lett. {\bf 102}, 206407 (2009).

\bibitem{mooden88}
A. R. Moodenbaugh, Y. Xu, M. Suenaga, T. J. Folkerts, R. N. Shelton,
 Phys. Rev. B {\bf 38}, 4596 (1988).

\bibitem{wu11}
T. Wu, H. Mayaffre, S. Kr\"{a}mer, M. Horvati\'{c}, C. Berthier,
 W. N. Hardy, R. Liang, D. A. Bonn, M. H. Julien, Nature {\bf 477}, 191
 (2011). 




\end{references}
\end{document}